\documentclass[pre,aps,onecolumn]{revtex4}
\usepackage{epsfig}
\usepackage{graphicx}
\usepackage{textcomp}

\begin{document}
\title{Large fluctuations in stochastic population dynamics: momentum space calculations}
\author{Michael Assaf$^1$, Baruch Meerson$^{1}$ and Pavel V Sasorov$^2$} \affiliation{$^1$Racah Institute of Physics, Hebrew University of
Jerusalem, Jerusalem 91904, Israel\\$^2$Institute of Theoretical and
Experimental Physics, Moscow 117218, Russia
\\E-mail: assaf@phys.huji.ac.il, meerson@cc.huji.ac.il, pavel.sasorov@gmail.com\\
URL: http://www.phys.huji.ac.il/$\sim$meerson/\hspace{1cm}}
\begin{abstract}
\large\textbf{Abstract.} \normalsize Momentum-space representation renders an interesting perspective to theory of large fluctuations in populations undergoing Markovian stochastic gain-loss processes. This representation is obtained when the master equation for the probability distribution of the population size is transformed
into an evolution equation for the probability generating
function. Spectral decomposition then brings about an eigenvalue problem for a non-Hermitian linear differential operator. The ground-state eigenmode encodes the stationary distribution of the population size. For long-lived metastable populations which exhibit extinction or escape to another metastable state, the quasi-stationary distribution and the mean time to extinction or escape are encoded by the eigenmode and eigenvalue of the lowest excited state. If the average population size in the stationary or quasi-stationary state is large, the corresponding eigenvalue problem can be solved via WKB approximation amended by other asymptotic methods.  We illustrate these ideas in several model examples.
\end{abstract}
\maketitle
\tableofcontents

\section{Introduction}
This work deals with dynamics of populations experiencing intrinsic noise caused by the discreteness of individuals and stochastic character of their interactions. When the average size $N$ of such a (stationary or quasi-stationary)  population is large, the noise-induced fluctuations in the observed number of individuals are typically small, and only rarely large.  In many applications, however, the rare large fluctuations can be very important. This is certainly true when their consequences are catastrophic, such as in the case of extinction of an isolated self-regulating
population after having maintained a long-lived metastable state, with applications ranging from population biology \cite{bartlett,assessment} and epidemiology \cite{bartlett,epidemic}
to genetic regulatory networks in living cells \cite{bio}. Another example of a catastrophic transition driven by a rare large intrinsic fluctuation is population explosion \cite{MS}.  Rare large fluctuations may also induce stochastic switches between different metastable states \cite{dykman}; these appear in genetic regulatory networks \cite{switches} and in other contexts.  Less dramatic but still important examples involve large fluctuations in the production rates of molecules on the surfaces of micron-sized dust grains in the interstellar medium, where the number of atoms, participating in the chemical reactions,  can be relatively small \cite{green,biham}.   As stochastic population dynamics is far from equilibrium and therefore
defies standard methods of equilibrium statistical mechanics,  large fluctuations of stochastic populations are of much interest to physics \cite{kampen,gardiner}.

In this paper we consider single-species populations which are well mixed, so that spatial degrees of freedom are irrelevant. To account for stochasticity of gain-loss processes (in the following -- reactions) and discreteness of the individuals (in the following -- particles), we assume a Markov process and employ a master equation which describes the time evolution of the probability $P_n(t)$ to have a population size $n$ at (continuous) time $t$. If the population exhibits neither extinction, nor a switch to another metastable state or to an infinite population size, a natural goal is to determine the stationary probability distribution of the population size \cite{kubo}. For metastable populations -- populations experiencing either extinction, or switches between different metastable states -- one is usually interested in the mean time to extinction or escape (MTE), and in the long-lived \textit{quasi}-stationary distribution (QSD) of the population size. For single-step
 processes these quantities  can be calculated by standard methods. The MTE can be calculated exactly by employing the backward master
equation \cite{kampen,gardiner}. This procedure yields an exact but unwieldy analytic expression for the MTE which, for a large population size in the metastable state, can be simplified via a saddle-point approximation \cite{Doering}. In its turn, the QSD of a single-step process can be found, in many cases, through a recursion.

For multi-step processes neither the MTE, nor QSD can be calculated exactly. Many practitioners have used, in different contexts of physics, chemistry, population biology, epidemiology, cell biology, \textit{etc}, what is often called ``the diffusion approximation": an approximation of the master equation by a Fokker-Planck equation. The latter can be
obtained via the van Kampen system size expansion, or other related prescriptions. With a Fokker-Planck equation at hand, the MTE and QSD can again be evaluated by standard methods \cite{kampen,gardiner}. Unfortunately, this approximation is in general uncontrolled, and fails in its description of
the tails of the QSD. As a result, it gives exponentially large errors in the MTE~\cite{Doering,gaveau,kessler,Assaf2}.

Until recently, the MTE and QSD had been calculated accurately only for a few model problems involving multi-step processes. Recently, Escudero and Kamenev \cite{EK} and Assaf and Meerson \cite{Assaf4} addressed quite a general set of reactions and developed controlled WKB approximations for the MTE and QSD for population switches \cite{EK} and population extinction \cite{Assaf4}. When necessary, the WKB approximation must be supplemented by a recursive solution of the master equation at small population sizes \cite{MS,Assaf4}, and by the van Kampen system size expansion in narrow regions where the ``fast" and ``slow" WKB modes are coupled and the WKB approximation fails~\cite{MS,EK,Assaf4}.

The techniques developed in Refs.~\cite{EK} and \cite{Assaf4} (see also Refs. \cite{dykman,kessler,MS}) were formulated directly in the space of population size $n$. An alternative approach invokes a complementary space which can be interpreted as a momentum space.
The momentum-space representation is obtained when  the master equation -- an infinite set of linear ordinary differential equations -- is transformed into a single evolution equation -- a linear partial differential equation -- for the probability generating function $G(p,t)$. Here $p$, a complementary variable, is conjugate to the population size $n$ and plays the role of ``momentum" in an effective Hamiltonian system which encodes, in the leading order of $1/N$-expansion, the stochastic population dynamics. One can then perform spectral decomposition of this linear partial differential equation for $G(p,t)$. In order to describe the stationary or metastable states, it suffices to consider the ground state and the lowest excited state of this spectral decomposition, whereas higher modes only contribute to short-time transients \cite{Assaf1,Assaf2}.

The ordinary differential equations for the ground state and the lowest excited state are determined by the specific set of reactions the population undergoes. The order of these equations is equal to the highest order of inter-particle reactions. For example, for two- (three-) body reactions the equations are of the second (third) order, \textit{etc}. In general, these ordinary differential equations cannot be solved exactly, and some perturbation techniques, employing the small parameter $1/N \ll 1$, need to be used.

The momentum-space spectral theory was developed~\cite{Assaf1,Assaf2,Assaf} for two-body reactions. Here we extend the theory to any many-body reactions. We also determine, in the general case,  the previously unknown boundary conditions for the above-mentioned eigenvalue problems. If there is no absorbing state at infinity, the boundary conditions are ``self-generated" by the demand that the probability generating function $G(p,t)$ be, at any $t$,  an entire function on the \textit{complex} plane $p$ \cite{entire}. We show that, for two-body reactions, the population extinction problem can always be solved by matching the exact solution of a quasi-stationary equation for the lowest excited state (see below) with a perturbative solution of a \textit{non}-quasi-stationary equation for the same state. This procedure always works when $N$ is sufficiently large. For three-, four-, $\dots$ body reactions the spectral decomposition can be used in conjunction with a $p$-space
WKB (Wentzel-Kramers-Brillouin) approximation which employs the same small parameter $1/N$ but does not rely on exact solution of the quasi-stationary equation. We find that there is a region of $p$ where the WKB approximation breaks down, and a region of $p$ where its accuracy is insufficient. In the former region a boundary-layer solution can be found and matched with the WKB solution. In the latter region a simple \textit{non}-WKB perturbative solution can be obtained.  The theory extensions presented here turn the  momentum-space spectral theory  of large fluctuations into a more general tool.

As the evolution equation for $G(p,t)$ is \textit{equivalent} to the master equation, the $p$-space approach is clearly advantageous, compared to the $n$-space approach, when the problem in the $p$ space admits an exact solution, see Refs.~\cite{green,gardiner}. Otherwise,
the technical advantages of the $p$-space approach are not \textit{a priori} obvious. In any case, it provides a viable alternative, and an interesting perspective, to theory of large fluctuations of stochastic populations.

Here is the layout of the rest of the paper. Section~II briefly introduces the  momentum-space spectral formalism, whereas in sections~III and IV we describe the methods of solution and  illustrate them on several model examples.  Sec.~III deals with a well-studied prototypical chemical reaction scheme which describes a stationary production of hydrogen molecules on interstellar dust grains. Here we show that the WKB approximation not only gives accurate results for the production rate (including its fluctuations) of hydrogen molecules, but also yields a complete stationary probability distribution function of the number of hydrogen atoms, including its non-Gaussian tails. In Sec.~IV we deal with isolated populations undergoing intrinsic-noise-driven extinction after maintaining a long-lived metastable state. Here, after some general arguments, we consider two different examples -- one studied previously and one new -- and determine the MTE and QSD. Throughout the paper we compare our analytical results with numerical solutions of the pertinent master equation and, when possible, with previous analytical results. Section~V summarizes our findings and discusses the advantages and disadvantages of the $p$-space method compared with the ``real" space WKB method \cite{dykman,kessler,MS,EK,Assaf4}.

\section{Master equation, probability generating function and spectral formulation}
Populations consist of discrete ``particles" undergoing stochastic gain and loss reactions. To account for both discreteness and stochasticity,
we assume the Markov property, see \textit{e.g.} Refs. \cite{kampen,gardiner}, and employ the master equation
\begin{equation}\label{master0}
\dot{P}_n(t)=\sum_{n^{\prime}\neq n} W_{n^{\prime} n} P_{n^{\prime}}
- W_{n n^{\prime}} P_{n}
\end{equation}
which describes the time evolution of  the probability distribution function $P_n(t)$ to
have $n$ particles at time $t$. Here $W_{n n^{\prime}}$ is the
transition rate matrix;  it is assumed that $P_{n<0}=0$.

The probability generating function, see \textit{e.g.} Refs. \cite{kampen,gardiner}, is defined as
\begin{equation}\label{genprob}
G(p,t)=\sum_{n=0}^{\infty} p^n P_n(t)\,.
\end{equation}
Here $p$ is an auxiliary variable which is conjugate to the number of particles $n$.
Once $G(p,t)$ is known, the probability distribution function $P_n(t)$ is given by the Taylor coefficients
\begin{equation}\label{prob}
P_n(t)=\left.\frac{1}{n!}\frac{\partial ^n G}{\partial
p^n}\right|_{p=0}
\end{equation}
or, alternatively, by employing the Cauchy theorem
\begin{equation}\label{cauchy}
P_n(t)=\frac{1}{2\pi i}\oint \frac{G(p,t)}{p^{n+1}}dp,
\end{equation}
where the integration has to be performed over a closed contour in
the \textit{complex} $p$-plane around the singular point $p=0$. For stochastic populations which do not exhibit
population explosion \cite{MS}, the probability $P_n(t)$ decays faster than exponentially at large $n$. Therefore, $G(p,t)$ is an entire function of $p$ on the complex
$p$-plane \cite{entire}.

If the reaction rates are polynomial in $n$, one can transform the master equation (\ref{master0}) into a single linear partial differential equation for the probability generating function,
\begin{equation}\label{geneq}
\frac{\partial G}{\partial t}= \hat{{\cal L}} G\,,
\end{equation}
where $\hat{{\cal L}}$ is a linear differential operator which includes powers of the partial differentiation operator $\partial/\partial p$. Equation~(\ref{geneq}) is exact and equivalent to the master equation (\ref{master0}). If only one-body reactions are present, $\hat{{\cal L}}$ is of first order in $\partial/\partial p$, and Eq.~(\ref{geneq}) can be solved by characteristics \cite{gardiner}. For many-body reactions one can proceed by expanding $G(p,t)$ in the yet unknown eigenmodes and eigenvalues of the problem \cite{Assaf,Assaf1,Assaf2}:
\begin{equation}\label{genexpansion}
G(p,t)= G_{st}(p)+\sum_{k=1}^{\infty} a_k \phi_{k}(p) e^{-E_{k}t}\,.
\end{equation}
As a result, partial differential equation~(\ref{geneq}) is transformed into an infinite set of ordinary differential equations: for the (stationary) ground-state  mode $G_{st}(p)$ and for the eigenmodes of excited states $\{\phi_k(p)\}_{k=1}^{\infty}$. By virtue of Eq.~(\ref{prob}) or (\ref{cauchy}), the ground state eigenmode determines the \textit{stationary} probability distribution function of the system. If a long-lived population ultimately goes extinct, the stationary distribution is trivial: $P_n=\delta_{n,0}$, where $\delta_{n,0}$ is the Kronecker's delta. What is of interest in this case is the \textit{quasi-stationary} distribution and its (exponentially long) decay time which yields an accurate approximation to the MTE. These quantities are determined by the lowest excited eigenmode $\phi(p)$ and the eigenvalue $E_1$, respectively  \cite{Assaf1,Assaf2}. The higher modes only contribute to short-time transients. Therefore, in the following we will focus on determining $G_{st}$ or solving th
 e eigenvalue problem for $\phi_1(p)$ and $E_1$.

\section{Stationary distributions: Ground-state calculations}
As a first example, we consider a simple model of production of $H_2$ molecules on micron-sized dust grains in interstellar medium. This model was investigated by Green \textit{et. al.} \cite{green}, who computed the stationary probability distribution function of the number of hydrogen atoms via finding an exact solution to the ordinary differential equation for $G_{st}$. The same results were obtained, by a different method, by Biham and Lipshtat \cite{biham}. We will use this problem as a benchmark of the ground-state calculations using the momentum-space WKB approach. As we will see, this approach gives, for $N\gg 1$, an accurate \textit{approximate} solution for $G_{st}(p)$, and so it can be employed for many other models where no exact solutions  are available.

Consider the following set of reactions: absorption of $H$-atoms by the  grain surface $\emptyset \stackrel{\alpha}{\rightarrow} H$, desorption of $H$-atoms, $H\stackrel{\beta}{\rightarrow} \emptyset$, and formation of $H_2$-molecules from pairs of $H$-atoms which can be formally described as annihilation $2H\stackrel{\gamma}{\rightarrow} \emptyset$.

\begin{figure}
\includegraphics[width=9.0cm,height=6.6cm,clip=]{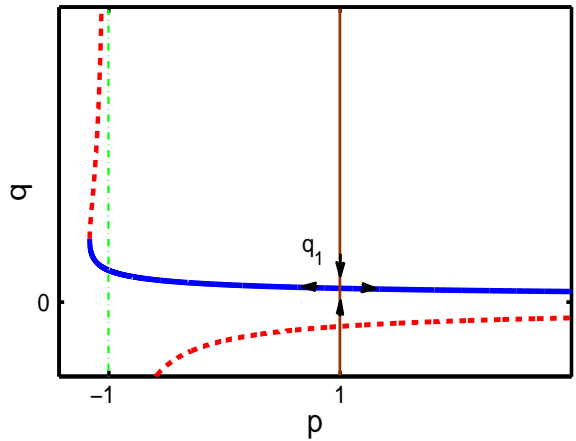}
\caption{Molecular hydrogen production on a grain. Shown are zero-energy
orbits of Hamiltonian (\ref{Ham0}) on the phase plane $(p,q)$. The
thick solid line corresponds to the instanton $q=q_+(p)$, see Eq.~(\ref{spr}). The motion along the vertical line $p=1$ is described by the deterministic rate equation ~(\ref{rateH}). The dashed lines depict the branch $q=q_-(p)$. It is non-physical at $q<0$ and does not contribute
to the WKB solution at $q>0$.} \label{phasehyd}
\end{figure}

To calculate the production rate of $H_2$-molecules, one needs to determine the stationary probability distribution function of the $H$-atoms,  $P_n(t\to\infty)$.  For convenience, we rescale time and reaction rates by the desorption rate $\beta$ and denote $N=2\beta/\gamma$ and $R=\alpha\gamma/(2\beta^2)$.  Ignoring fluctuations, one can write down the following (rescaled) deterministic rate equation:
\begin{equation}
\dot{\bar{n}}=NR-\bar{n}-\frac{2}{N}\bar{n}^2\,,
\label{rateH}
\end{equation}
where $\bar{n}(t) \gg 1$ is the average population size. The only positive fixed point of this equation,
\begin{equation}\label{rateeq1}
\bar{n}=\frac{N}{4}(\sqrt{1+8R}-1),
\end{equation}
is attracting, and the stationary probability distribution function $P_n$ is expected to be peaked around it. The master equation describing the stochastic dynamics of this system in rescaled time is
\begin{eqnarray}
\label{master1}
\frac{d}{dt}{P}_{n}(t)=\frac{1}{N}\left[(n+2)(n+1)P_{n+2}(t)-n(n-1)P_{n}(t)\right]+\left[(n+1)P_{n+1}(t)-nP_{n}(t)\right]+NR(P_{n-1}-  P_n)\,.
\end{eqnarray}
This yields the following partial differential equation for $G(p,t)$ \cite{green}:
\begin{equation}\label{pde1}
\frac{\partial G}{\partial t} =
\frac{1}{N}(1-p^2)\frac{\partial^2 G}{\partial p^2}+(1-p)\frac{\partial G}{\partial p}+NR(p-1)G\,.
\end{equation}
The steady-state solution $G_{st}$ obeys the ordinary differential equation
\begin{equation}\label{ode1}
\frac{1}{N}(1+p)G_{st}^{\prime\prime}+G_{st}^{\prime}-NRG_{st}=0\,,
\end{equation}
where primes denote the $p$-derivatives. The boundary conditions are ``self-generated". Indeed, equality $G (p=1,t)=1$ holds at all times. This reflects conservation of probability, see Eq.~(\ref{genprob}). Therefore,
\begin{equation}\label{at1}
    G_{st}(1)=1\,.
\end{equation}
Furthermore, Eq.~(\ref{ode1}) has a singular point at $p=-1$. As $G_{st}(p)$ must be analytic at $p=-1$,
we demand
\begin{equation}\label{secbc}
G_{st}^{\prime}(-1)-NRG_{st}(-1)=0.
\end{equation}
The boundary-value problem (\ref{ode1})-(\ref{secbc}) is exactly solvable in special functions \cite{green}. For a general set of reactions, however, one cannot expect an exact solution. Still, one can employ the small parameter $1/N$ to develop an accurate analytical approximation. To illustrate this point we will proceed as we were unaware of the exact solution, and then compare the approximate solution with the exact one. As the small parameter $1/N$ appears in the coefficient of the highest derivative, it is natural to use (a dissipative variant of) the stationary WKB approximation in the $p$-space \cite{Assaf}. The WKB ansatz is
\begin{equation}\label{ansatz}
G_{st}(p)=a(p)e^{-NS(p)},
\end{equation}
where the action $S(p)$ and amplitude $a(p)$ are non-negative functions of $p$. Using this ansatz in Eq.~(\ref{pde1}) with a zero left hand side, we obtain
\begin{eqnarray}\label{wkbfull1}
\frac{1}{N}(1-p^2)\left[a^{\prime\prime}-2NS^{\prime}a^{\prime}-NS^{\prime\prime}a+N^2(S^{\prime})^2 a\right]+(1-p)(a^{\prime}-NS^{\prime}a-N R\,a)=0\,.
\end{eqnarray}
In the leading order ${\cal O}(N)$ we obtain a stationary Hamilton-Jacobi equation
$H[p,-S^{\prime}(p)]=0$ with zero energy, \textit{cf}. Ref. \cite{Kamenev1}. The effective Hamiltonian is
\begin{equation}\label{Ham0}
H(p,q)=(1-p)[(1+p)q^2+q-R],
\end{equation}
where we have introduced $q(p)=-S^{\prime}(p)$: the reaction coordinate conjugate to the momentum $p$. The trivial zero-energy phase orbit $p=1$ is an invariant line of the Hamiltonian; it  corresponds to the deterministic dynamics \cite{Kamenev1}. Indeed, the Hamilton's equation for $\dot{q}$,
$$
\dot{q}=R-q-2 q^2,
$$
coincides, in view of the relation $q=n/N$, with the deterministic rate equation~(\ref{rateH}).  Hamiltonian (\ref{Ham0}) also has two nontrivial invariant zero-energy lines which are composed of the two solutions, $q_-(p)$ and $q_+(p)$, of
the quadratic equation $(1+p)q^2+q-R=0$:
\begin{equation}\label{spr}
q_-(p)=\frac{-1-v(p)}{2(1+p)}\,,\;\;\;q_+(p)=\frac{-1+v(p)}{2(1+p)}\,.
\end{equation}
Here we have denoted
\begin{equation}\label{vp}
v(p)=\sqrt{1+4R(1+p)}.
\end{equation}
The phase plane of this system is shown in Fig.~\ref{phasehyd}. The phase orbits $q=q_{-}(p)$ must be discarded. This is because $q_{-}(p)$ diverges at $p=-1$, whereas $G_{st}(p)$, and therefore $S(p)$, must be analytic everywhere.

The remaining nontrivial zero-energy phase orbit $q_+(p)\equiv q(p)$ has a special role. It describes the most probable path along which the system evolves, (almost) with certainty, in the course of a fluctuation bringing the system from the fixed point $(1,q_1)$ in the phase space $(p,q)$ to a given point, see Fig.~\ref{phasehyd}. Here $q_1=(1/4)(\sqrt{1+8R}-1)]$ is the attracting point of the deterministic rate equation, see Eq.~(\ref{rateeq1}).

Integrating the equation $S^{\prime}(p)=-q_+(p)$, we obtain
\begin{eqnarray}\label{act1}
S(p)=-v(p)+v(1)+\ln\frac{v(p)+1}{v(1)+1}\,,
\end{eqnarray}
where we have fixed the definitions of $a(p)$ and $S(p)$ by demanding
$S(p=1)=0$.

To calculate the amplitude $a(p)$ we proceed to the subleading ${\cal O}(1)$ order in Eq.~(\ref{wkbfull1}):
\begin{equation}
-2(1+p)S^{\prime}a^{\prime}-(1+p)S^{\prime\prime}a+a^{\prime}=0.
\end{equation}
Using $S(p)$ from Eq.~(\ref{act1}), we arrive at a first-order ordinary differential equation for $a(p)$,
\begin{equation}
\frac{a^{\prime}(p)}{a(p)}=\frac{4R^2(1+p)}{v(p)^2[1+v(p)]^2}\,.
\end{equation}
Solving this equation, we obtain the WKB solution
\begin{equation}\label{gwkb}
G_{st}^{WKB}(p)=\frac{v(1)^{1/2}\left[1+v(p)\right]}{v(p)^{1/2}[1+v(1)]}e^{-NS(p)},
\end{equation}
where the integration constant is chosen so as to obey boundary condition~(\ref{at1}).  As one can easily check, WKB solution (\ref{gwkb}) also obeys boundary condition (\ref{secbc}).

As expected, pre-exponent $a(p)$ of the WKB solution (\ref{gwkb}) diverges at the turning point $p_{tp}=-1-1/(4R)<-1$ of the zero-energy phase orbit, see Fig. ~\ref{phasehyd}. As a result, the WKB solution breaks down in a close vicinity of this point. At $p<p_{tp}\,$ a WKB solution of a different nature appears:  it exhibits decaying oscillations as a function of $p$. The oscillating WKB solution can be found by treating $S(p)$ as a complex-valued, rather than real, function. We will not need the oscillating solution, because the non-oscillating one, Eq.~(\ref{gwkb}), turns out to be sufficient for the purpose of calculating the probabilities $P_n$, see below.

\begin{figure}
\includegraphics[width=9.0cm,height=6.6cm,clip=]{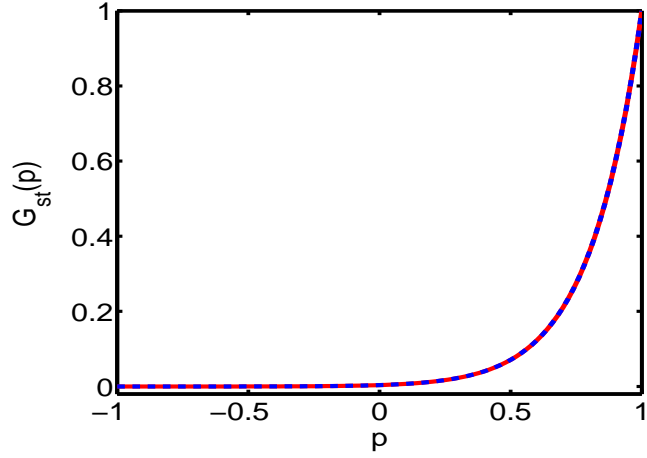}
\caption{Molecular hydrogen production on a grain. Shown is a comparison of WKB result (\ref{gwkb}) for $G_{st}(p)$ (dashed line) and exact result  (\ref{gex}) (solid line) for $N=10$ and $R=1$. The agreement is excellent even for this moderate $N$.} \label{gcomp}
\end{figure}

Now we can compare  WKB solution (\ref{gwkb}) with the exact solution of the problem (\ref{ode1})-(\ref{secbc}), derived by Green \textit{et al.} \cite{green}:
\begin{equation}\label{gex}
G_{st}^{exact}(p)=\left(\frac{2}{1+p}\right)^{\frac{N-1}{2}}\frac{I_{N-1}[2N \sqrt{R(1+p)}]}{I_{N-1}(2N \sqrt{2R})}\,,
\end{equation}
where
$I_k(w)$ is the modified Bessel function. To this end let us calculate the large-$N$ asymptote of $I_{N-1}[2N \sqrt{R(1+p)}]$ by using the integral definition of the modified Bessel function \cite{Abramowitz}
\begin{eqnarray}
I_{N-1}[2N\sqrt{R(1+p)}]=\frac{\left[N^2 R(1+p)\right]^{\frac{N-1}{2}}}{\sqrt{\pi}\,\Gamma(N-1/2)}\int_{-1}^1\frac{(1-t^2)^N e^{-2N \sqrt{R(1+p)} t}}{(1-t^2)^{3/2}}dt,
\end{eqnarray}
where $\Gamma(\dots)$ is the Euler Gamma function.
As $N\gg 1$, we can evaluate the integral by the saddle point approximation \cite{orszag}. Denoting $f(t)=\ln (1-t^2)-2\sqrt{R(1+p)} t,$
we find the relevant saddle point
$$
t_*(p)=\frac{1-\sqrt{1+4R(1+p)}}{2\sqrt{R(1+p)}}=-\sqrt{\frac{v(p)-1}{v(p)+1}}\,,
$$
with $v(p)$ from Eq.~(\ref{vp}).  Then, expanding $f(t)\simeq f(t_*)+(1/2)f^{\prime\prime}(t_*)(t-t_*)^2$ with $f^{\prime\prime}(t_*)=-v(p)[1+v(p)]$, and performing the Gaussian integration, we obtain the $N\gg 1$  asymptote
\begin{eqnarray}\label{fp}
I_{N-1}[2N\sqrt{R(1+p)}]\simeq \frac{1+v(p)}{2\sqrt{2}\,\Gamma(N-1/2)\sqrt{N v(p)}}\left[N^2 R(1+p)\right]^{\frac{N-1}{2}}e^{N\{v(p)-1+\ln 2-\ln[1+v(p)]\}}\,.
\end{eqnarray}
Note that the saddle point approximation is valid on the entire segment $-1\leq p\leq 1$. In particular,  Eq.~(\ref{fp}) with $p=1$ yields the $N\gg 1$ asymptote of the denominator of Eq.~(\ref{gex}). Now one can see that the large-$N$ asymptote of Eq.~(\ref{gex}) exactly coincides with WKB solution (\ref{gwkb}). Actually, the WKB result is indistinguishable from the exact result already for $N=10$, see Fig.~\ref{gcomp}.

Of a primary interest in the context of astrochemistry is the mean and variance of the steady-state production rate of $H_2$ molecules. Going back to physical units, we can write the mean steady-state production rate as
$$
{\cal R}(H_2)=\frac{\gamma}{2}\displaystyle\sum_{n=0}^{\infty}n (n-1) P_n = \frac{\gamma}{2} \langle n(n-1)\rangle=\frac{\gamma}{2}G_{st}^{\prime\prime}(1),
$$
\begin{equation}
{\cal R}(H_2)\simeq \frac{2\gamma N^2 R^2}{[v(1)+1]^2}\left[1-\frac{1}{N v^2(1)}\right]\,,
\label{prodrate}
\end{equation}
where $v(p)$ is given by Eq.~(\ref{vp}). One can check that this expression coincides with that obtained from the exact result, see Eq.~(22) in Ref.~\cite{green}, in the leading- and subleading-order at $N\gg 1$.  The leading term in Eq.~(\ref{prodrate})  is what the deterministic
rate equation (\ref{rateH}) predicts.

Now consider the variance of the steady-state production rate of $H_2$ molecules:
$$
V(H_2)=\frac{\gamma}{2} \left[\langle n^2 (n-1)^2\rangle - \langle n (n-1)\rangle^2\right]\,.
$$
Using identity
\begin{eqnarray}
 n^{2}(n-1)^{2} =  n(n-1)(n-2)(n-3) + 4n(n-1)(n-2) + 2n(n-1)\,, \nonumber
\end{eqnarray}
we obtain the exact relation
$$
V(H_2) =\frac{\gamma}{2}\left\{G^{IV}(1)+4G^{\prime\prime\prime}(1)+
2G^{\prime\prime}(1)-[G^{\prime\prime}(1)]^2\right\}.
$$
From WKB solution (\ref{gwkb}) we obtain in the leading order
\begin{equation}
V(H_2)\simeq \frac{16 \gamma N^3 R^3 \left[v(1)+6R+1\right]}{v(1) \left[v(1)+1\right]^4}\,.
\end{equation}
The relative fluctuations of the production rate, $\sqrt{V}/{\cal R}$, scale with $N$ as $N^{-1/2}$, as expected.

\begin{figure}
\includegraphics[width=9.0cm,height=6.6cm,clip=]{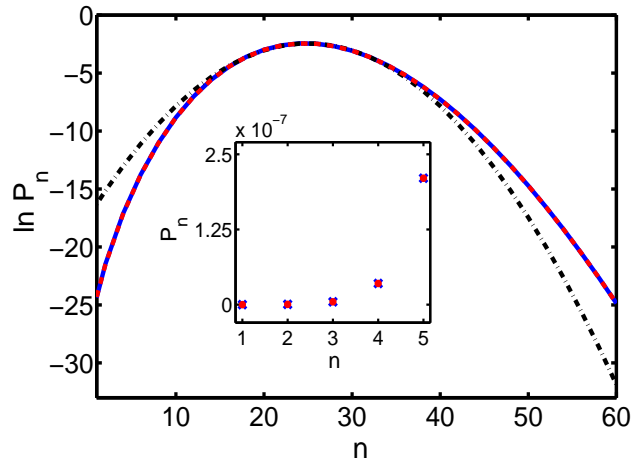}
\caption{Molecular hydrogen production on a grain. Shown is the natural logarithm of the stationary distribution $P_n$ versus $n$ for $N=50$ and $R=1$. The solid line is WKB approximation (\ref{pdfap}), the dashed line is
exact solution (\ref{pdfex}), and the dash-dotted line is the Gaussian
approximation. The WKB approximation and the exact solution are indistinguishable for all $n$. The non-Gaussian tails of the distribution cannot be described correctly by the van Kampen system size expansion. The inset shows, by different symbols, the small-$n$ asymptote of the distribution obtained analytically and numerically.}
\label{probhyd}
\end{figure}

Actually, the WKB approximation yields the whole stationary probability distribution function of the number of $H$ atoms. Green \textit{et. al.} \cite{green} obtained this distribution exactly from Eqs.~(\ref{gex}) and (\ref{prob}):
\begin{equation}\label{pdfex}
P_n=2^{\frac{N-1}{2}}\frac{(N^2
R)^{n/2}}{n!}\frac{I_{N+n-1}(2N\sqrt{R})}{I_{N-1}(2N\sqrt{2R})}\,.
\end{equation}
The $N\gg 1$, $n\gg 1$ asymptote of (\ref{pdfex}) can be written as
\begin{eqnarray}\label{pdfap}
P_n\simeq \frac{\sqrt{(1+q)\,v(1)}}{\sqrt{2\pi q N\,u(q)}}\,\frac{1+u(q)}{1+v(1)}e^{N\left\{\ln[1+v(1)]-v(1)+q+(1+q)u(q)-\ln
[(1+q)(1+u(q))]-q\ln[q(1+q)(1+u(q))/(2R)]\right\}},
\end{eqnarray}
where $q=n/N$, $v(p)$ is given by Eq.~(\ref{vp}) and
\begin{equation}\label{un}
u(q)=\sqrt{1+4R/(1+q)^2}\,.
\end{equation}
Now we compare Eq.~(\ref{pdfap}) with the WKB result, obtained from
Eqs.~(\ref{cauchy}) and (\ref{gwkb}):
\begin{equation}\label{pdfwkb}
P_n^{WKB}=\frac{1}{2\pi i}\oint dp \frac{v(1)^{1/2}\left[1+v(p)\right]}{p
\,v(p)^{1/2}[1+v(1)]}e^{-NS(p)-n\ln p}\,,
\end{equation}
where $S(p)$ is given by Eq.~(\ref{act1}). As $n\gg 1$, we can evaluate the
integral via the saddle point approximation. Let
$f(p)=-NS(p)-n\ln p$. The saddle point is at
$p_*=q(1+q)[1+u(q)]/(2R)$, where $u(q)$ is given by Eq.~(\ref{un}).
As $f^{\prime\prime}(p_*)>0$, the
integration contour in the vicinity of the saddle point must be chosen
perpendicular to the real axis. This adds an additional phase of
$e^{i\pi/2}$ to the
solution \cite{orszag}, which cancels $i$ in the denominator
of Eq.~(\ref{pdfwkb}). After the Gaussian integration and some algebra Eq.~(\ref{pdfwkb}) coincides with Eq.~(\ref{pdfap}). Finally, one can calculate $P_n$ at $N\gg 1$ but $n={\cal O}(1)$ by directly differentiating the WKB result (\ref{gwkb}) for $G_{st}(p)$, see Eq.~(\ref{prob}). The resulting probability distribution function is shown in Fig.~\ref{probhyd}. As one can see, the agreement between the WKB distribution and the exact distribution is excellent for all $n$.

\section{Metastability and extinction: First-excited-state calculations}\label{sec2}
Now we switch to isolated stochastic populations, so that there is no influx of particles into the system. If there is no population explosion, isolated populations ultimately undergo extinction with probability one. The deterministic rate equation for such a population can be written as
\begin{equation}\label{genrateeq}
\dot{\bar{n}}=\bar{n}\psi(\bar{n}),
\end{equation}
where $\psi(\bar{n})$ is a smooth function. In the following we assume $\psi(0)>0$, so that $\bar{n}=0$ is a \textit{repelling} fixed point of Eq.~(\ref{genrateeq}). The deterministically stable population size corresponds to an \textit{attracting} fixed point $\bar{n}=n_1>0$.  According to the classification of Ref.~\cite{Assaf4}, such populations exhibit scenario A of extinction.

Let $n_1={\cal O}(N)\gg 1$. After a short relaxation time $t_r$, the population typically converges into a long-lived metastable state whose population size distribution is peaked around $n=n_1$. This metastable probability distribution function is encoded in the lowest excited eigenmode $\phi(p)\equiv\phi_1(p)$ of the probability generating function $G(p,t)$ (\ref{genexpansion}). Indeed, at $t\gg t_r$, the higher eigenmodes in the spectral expansion (\ref{genexpansion}) have already decayed, and $G(p,t)$ can be approximated as \cite{Assaf1,Assaf2}
\begin{equation}\label{G2}
G(p,t)\simeq 1-\phi(p)e^{-E t},
\end{equation}
where the lowest excited eigenfunction is normalized so that $\phi(0)=1$. The (exponentially small) lowest excited eigenvalue $E\equiv E_1$ determines the MTE of the population, $E\simeq\tau_{ex}^{-1}$. The slowly time-dependent probability distribution function of the population size, at $t\gg t_r$, is
\begin{equation}\label{qsd}
P_{n>0}(t)\simeq \pi_n e^{-t/\tau_{ex}}\;,\;\;P_0(t)\simeq 1-e^{-t/\tau_{ex}}\,.
\end{equation}
That is, the metastable probability distribution function exponentially slowly decays in time, whereas the extinction probability $P_0(t)$ exponentially slowly grows and reaches $1$ at $t\to \infty$. The shape function $\pi_n$ of the metastable distribution is called the quasi-stationary distribution (QSD).  The QSD and MTE of a metastable population can be obtained by solving the eigenvalue problem for $\phi(p)$ and $E$, respectively. We now discuss some general properties of the solution to this eigenvalue problem, whereas in the following subsections we will illustrate the method of solution on two examples.

\subsection{General considerations}
Plugging Eq.~(\ref{G2}) into Eq.~(\ref{geneq}), we arrive at an ordinary differential equation for $\phi(p)$:
\begin{equation}\label{odeeq}
\hat{{\cal L}} \phi+E\phi=0\,.
\end{equation}
As $G(p,t)$ is an entire function on the complex $p$-plane \cite{entire}, $\phi(p)$ must be analytic in all singular points of differential operator $\hat{{\cal L}}$. If the order of this operator is $K$, this demand yields $K$ ``self-generated" boundary conditions for $\phi(p)$. In view of the equality $G(p=1,t)=1$, operator $\hat{{\cal L}}$ vanishes at $p=1$, which yields a universal boundary condition: $\phi(1)=0$. The rest of the $K-1$ self-generated boundary conditions are problem-specific, see examples below.

\begin{figure}
\includegraphics[width=9.0cm,height=6.6cm,clip=]{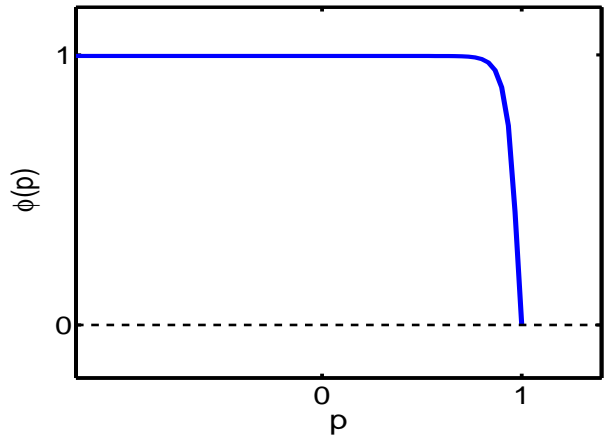}
\caption{Shown is a sketch of the eigenfunction $\phi(p)$ of the lowest excited state at $N\gg 1$ for a typical problem of population extinction. $\phi(p)$
is almost constant on the region $p<1$ except close to $p=1$, where it rapidly goes to
zero.} \label{phipic}
\end{figure}

What is the general structure of differential operator $\hat{{\cal L}}$? For populations that experience extinction, $\hat{{\cal L}}\phi$ cannot include a term proportional to $\phi$, as such a term would correspond to influx of particles into the system, $\emptyset\to A$, and would prevent extinction. In general, $\hat{{\cal L}}$ includes first-order derivative terms (corresponding to branching and decay processes) and higher-order derivative terms. For extinction scenario A one has $\psi(0)>0$, see Eq.~(\ref{genrateeq}). Let  $b_0$ denote the rate of decay $A\to\emptyset$, and $b_m$, $m=2,3, \dots, M,$ denote the rates of  branching reactions $A\to mA$. One has $\psi(0)\equiv b_2+2b_3+\dots+(M-1)b_M-b_0>0$. Rescaling time by $\psi(0)$, we see that the (rescaled) coefficient of the term $\bar{n}^j$ (for $j=1,2,\dots$) in Eq.~(\ref{genrateeq}) must scale as $N^{1-j}$ to ensure that $n_1={\cal O}(N)$. As a result, the (rescaled) coefficient of the $j$th-order derivative term in $\hat{{\cal L}}$ scales as $N^{1-j}$, and $\hat{{\cal L}}$ can be written as
\begin{equation}\label{L}
    \hat{{\cal L}}=f_1(p) \frac{d}{dp}+\frac{1}{N}\,f_2(p)\frac{d^2}{dp^2}+ \dots
    +\frac{1}{N^{K-1}}\,f_K(p)\frac{d^{K}}{dp^{K}}\,.
\end{equation}
For reaction rates that are polynomial in $n$, the functions $f_j(p)$ are polynomial in $p$. Notably, all functions $f_j(p)$ vanish at $p=1$.  How does the solution of Eq.~(\ref{odeeq}) look like at $N\gg 1$? As $E$ turns out to be exponentially small in $N$, the simplest approximation for Eq.~(\ref{odeeq}) would be to discard all terms except $f_1(p) d\phi/dp$, arriving at a constant solution $\phi(p)=1$ (according to our choice of normalization). Indeed, as $n_1={\cal O}(N)\gg 1$, the probability to observe $n\ll n_1$ particles in the metastable state is exponentially small. These probabilities are proportional to low-order derivatives of $\phi$ at $p=0$, see Eqs.~(\ref{prob}) and (\ref{G2}), so $\phi(p)$ must indeed be almost constant there. This solution, however, does not obey the zero boundary condition at $p=1$. The true solution, therefore, must rapidly fall to $0$ in a close vicinity of $p=1$, see Fig. \ref{phipic}.  The point $p=1$ is a singular point of Eq.~(\ref{odeeq}). Actually, when approaching $p=1$ from the left, the almost constant solution breaks down even earlier: in the vicinity of another point $p=p_f<1$ where $f_1(p)$ vanishes, see the next paragraph. In the vicinity of $p=p_f$, the first-order derivative term seizes to be dominant, and all terms in Eq.~(\ref{odeeq}), including $E\phi$, are comparable. Although $\phi(p)$ deviates from a constant value in the vicinity of $p=p_f$, one can still treat this deviation perturbatively: $\phi(p)\simeq 1+\delta\phi(p)$, where $\delta\phi\ll 1$. When $p$ becomes distinctly larger than $p_f$, $\phi(p)$ already varies strongly. Here, the $E\phi$-term [which comes from the time derivative of $G(p,t)$] can again be neglected, and so the (nontrivial) solution which is sought in this region is \textit{quasi-stationary}. The quasi-stationary solution can be found in the WKB approximation, as the typical length scale $1/N$, over which $\phi(p)$ varies, is much smaller here than  $1-p_f$ (a more accurate criterion will appear later).

Why does the root $p_f$ of function $f_1(p)$ exist? After some algebra, function $f_1(p)$ can be written as
\begin{equation}\label{f1p}
f_1(p)=\sum_{m=0}^M \tilde{b}_m (p^m-p)\,,
\end{equation}
where $\tilde{b}_m=b_m/\psi(0)$. The polynomial equation $f_1(p)=0$ has appeared in the context of $n$-space description of stochastic population extinction~\cite{Assaf4}. It has been shown in Ref.~\cite{Assaf4} that this equation
has exactly two real roots: $p=1$ and $p=p_f$, where in general $0\leq p_f<1$.

Now we can summarize the general scheme of solution of the eigenvalue problem for the lowest excited state.
One has to consider \textit{three} separate regions: (i) the region to the left, and sufficiently far, from the point $p=p_f$, where one can put $\phi(p)=1$ up to exponentially small corrections, (ii) in the boundary-layer region $|p-p_f|\ll p_f$ where $\phi(p)$ is still very close to $1$ and can be sought perturbatively, and (iii) in the quasi-stationary region $p_f<p\leq 1$, where $\phi(p)$ varies strongly, and the WKB approximation can be used. The solutions in the neighboring regions can be matched in their joint regions of validity. This procedure holds, at $N\gg 1$, for a broad class of systems exhibiting extinction.

There is a convenient shortcut to this general procedure when the highest-order reaction in the problem is two-body. Here the quasi-stationary equation [Eq.~(\ref{odeeq}) with the $E\phi$ term neglected] is always solvable exactly. There is no need to apply WKB approximation in such cases, and it suffices to only consider two, rather than three, regions, see the next subsection. Finally, regardless of the order of $\hat{{\cal L}}$, it is simpler to deal with $u(p)\equiv\phi^{\prime}(p)$ rather than with $\phi(p)$ itself, as this enables one to reduce the order of the ordinary differential equation by one everywhere.

\subsection{Branching-annihilation-decay}
The first example deals with a population of ``particles" which undergoes three stochastic reactions: branching
$A\stackrel{\lambda}{\rightarrow} 2A$, decay
$A\stackrel{\mu}{\rightarrow} \emptyset$ and annihilation
$2A\stackrel{\sigma}{\rightarrow} \emptyset$. As the state $n=0$ is absorbing, the population ultimately goes extinct. This example was solved by Kessler and Shnerb \cite{kessler} via ``real-space" WKB approximation, where the calculations are done in the space of population size. Here we solve it in the momentum space.  Because of the presence of the linear decay reaction $A \rightarrow 0$, this example exhibits a generic transcritical bifurcation as a function of the control parameter $R_0$ introduced below, and generalizes simple single-parameter models \cite{Assaf1,Assaf2} considered earlier. The deterministic rate equation reads
\begin{equation}\label{rateeq}
\dot{\bar{n}} = (\lambda-\mu) \bar{n}-\sigma\,\bar{n}^2\,.
\end{equation}
For $\lambda>\mu$ Eq.~(\ref{rateeq}) has, in addition to the trivial fixed point $\bar{n}=0$, also a positive fixed point $n_1= (\lambda-\mu)/\sigma$.  When starting from any
$\bar{n}(t=0) > 0$, the population size flows to the attracting
fixed point $\bar{n}=n_1$, with characteristic relaxation time
$t_r=(\lambda-\mu)^{-1}$, and stays there forever. Rescaling time $\lambda t \to t$, and introducing rescaled parameters, $N=\lambda/\sigma$ and
$R_0=\lambda/\mu$, the attracting fixed point becomes $n_1=N(1-R_0^{-1})$. We demand that $N\gg 1$, and $R_0>1$ and not too close to $1$ (the exact criterion will appear later). When $R_0$ exceeds $1$ the deterministic system undergoes a transcritical bifurcation.

To account for intrinsic noise we consider the master equation
\begin{eqnarray}
\label{p10}
\frac{d}{dt}{P}_{n}(t)=\frac{1}{2N}\left[(n+2)(n+1)P_{n+2}(t)-n(n-1)P_{n}(t)\right]+(n-1)P_{n-1}(t)-nP_{n}(t)+\frac{1}{R_0}\left[(n+1) P_{n+1}- n P_n\right]\,,
\end{eqnarray}
where time is rescaled, $\lambda t \to t$. The evolution equation for the probability generating function $G(p,t)$ is
\begin{equation}\label{pde2}
\frac{\partial G}{\partial t}=\frac{1}{2N}(1-p^2)\frac{\partial^2
G}{\partial p^2}+(p-1)\left(p-\frac{1}{R_0}\right)\frac{\partial
G}{\partial p}\,.
\end{equation}
At  $t\gg t_r=(1-1/R_0)^{-1}$ the metastable probability distribution function, peaked at
$n\simeq n_1$,  sets in, and Eq.~(\ref{G2}) holds.
To determine the QSD and MTE we turn to the Sturm-Liouville problem for the lowest excited eigenmode $\phi(p)$ and eigenvalue $E$
\begin{equation}\label{ode2}
\frac{1}{2N}(1-p^2)\phi^{\prime\prime}+(p-1)\left(p-\frac{1}{R_0}\right)\phi^{\prime}+E\phi=0\,.
\end{equation}
Here, the self-generated boundary conditions for $\phi(p)$ are: $\phi(1)=0$ and $2(1+R_0^{-1})\phi^{\prime}(-1)+E\phi(-1)=0$. Because of the expected exponential smallness of $E$, the latter condition can be safely approximated by $\phi^{\prime}(-1)\simeq 0$.

We now apply the procedure of solution presented in the previous subsection on Eq.~(\ref{ode2}). Using $u(p)= \phi^{\prime}(p)$, the exact solution of the quasi-stationary equation [Eq.~(\ref{ode2}) without the $E\phi$ term],
\begin{equation}\label{ode2wkb}
\frac{1}{2N}(1-p^2)u^{\prime}+(p-1)\left(p-\frac{1}{R_0}\right)u=0\,,
\end{equation}
can be written as
\begin{equation}\label{phiwkb0}
u (p)=C e^{-NS(p)}.
\end{equation}
Here
\begin{equation}\label{S2}
S(p)=2\left[1-p+\left(1+\frac{1}{R_0}\right)\ln
\left(\frac{1+p}{2}\right)\right]\,.
\end{equation}
To determine the arbitrary constant $C$ we need a boundary condition for $u(p)$ at $p=1$.
It follows from Eq.~(\ref{G2}) that, at $t \gg t_r$,
\begin{equation}\label{dGdp}
\frac{\partial G}{\partial p}(1,t)\simeq -u(1) e^{-Et}\,.
\end{equation}
On the other hand, by virtue of Eq.~(\ref{genprob}), the left hand side of Eq.~(\ref{dGdp})
is equal to $\bar{n}(t)$ which behaves as $n_1\, \exp(-Et)$, see \textit{e.g.} Ref.~\cite{Assaf2}.  As a result, $u(1)\simeq-n_1$ and,
by using Eq.~(\ref{phiwkb0}), we obtain $C=-N(1-R_0^{-1})$. Therefore,
\begin{equation}\label{phiwkb2}
u(p)=-N\left(1-\frac{1}{R_0}\right)e^{-NS(p)}
\end{equation}
with $S(p)$ from Eq.~(\ref{S2}). This yields the solution we looked for: $\phi=\int_1^{p}u(s)ds$, which satisfies the boundary condition $\phi(1)=0$. One can check now that neglecting
the $E\phi$ term  in Eq.~(\ref{ode2}) demands $pR_0-1\gg N^{-1/2}$.

\begin{figure}
\includegraphics[width=9.0cm,height=6.6cm,clip=]{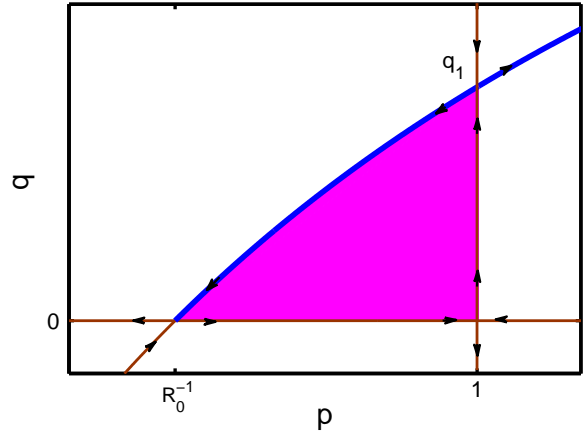}
\caption{Branching-annihilation-decay. Shown are zero-energy lines of Hamiltonian (\ref{Ham1}) on the $(p,q)$ phase plane. The
thick solid line corresponds to the instanton $q=-S^{\prime}(p)$ (\ref{S2}). Here $q_1=n_1/N=1-R_0^{-1}$, and the
area of the shaded region is equal to $S_0$ from
Eq.~(\ref{S0}).} \label{phasedb}
\end{figure}

Although there is no need in the WKB approximation in this case of a two-body reaction, it is still instructive to re-derive Eq.~(\ref{phiwkb2}) by using the WKB approximation for $\phi(p)$. To this end we consider the quasi-stationary version of Eq.~(\ref{ode2}),
\begin{equation}\label{ode2qsd}
\frac{1}{2N}(1-p^2)\phi^{\prime\prime}+(p-1)\left(p-\frac{1}{R_0}\right)
\phi^{\prime}=0\,,
\end{equation}
and make a WKB ansatz  $\phi(p)= a(p) \exp[-N S(p)]$. In the leading order in $N\gg1$ we obtain a stationary Hamilton-Jacobi equation
$H[p,-S^{\prime}(p)]=0$ with effective Hamiltonian \cite{Kamenev2}
\begin{equation}\label{Ham1}
H(p,q)=\left[p-\frac{1}{R_0}-\frac{(1+p)q}{2}\right]q(p-1).
\end{equation}
Here, as in Sec. III, $q(p)=-S^{\prime}(p)$ is the reaction coordinate conjugate to the momentum $p$. There are two trivial zero energy orbits of this Hamiltonian:  the deterministic orbit $p=1$ and the ``extinction orbit" $q=0$. The action along the extinction orbit  is zero: $S(p)=0$, so the corresponding WKB mode can be called ``slow".  There is also a nontrivial zero-energy orbit $q(p)=2(p-R_0^{-1})/(1+p)$. It includes a heteroclinic orbit exiting, at $t=-\infty$, the fixed point $(p=1,q=q_1\equiv n_1/N)$ and entering, at $t=\infty$, the fixed point $(p=R_0^{-1},q=0)$ of the phase plane $(p,q)$, see Fig. \ref{phasedb}. This orbit is the ``extinction instanton" \cite{Kamenev2,Kamenev1}. It describes the most probable path of the system from the long-lived metastable state to extinction. Integrating along this orbit and choosing $S(p=1)=0$, we recover Eq.~(\ref{S2}). This solution can be called the ``fast" WKB mode.

In the subleading order of the WKB approximation one obtains $a(p)=\left(1-R_0^{-1}\right)(1+p)/\left[2\left(p-R_0^{-1}\right)\right]$ for the fast, and $a(p)=const$ for the slow WKB modes. The general WKB solution is a superposition of the two modes,
\begin{equation}\label{alternative}
    \phi(p)=1-\frac{\left(1-R_0^{-1}\right)(1+p)}{2\left(p-R_0^{-1}\right)}e^{-N S(p)}\,,
\end{equation}
with $S(p)$ from Eq.~(\ref{S2}). Here we have already imposed the boundary condition $\phi(1)=0$ and normalization condition $\phi(0)\simeq 1$. The $p$-derivative of $\phi(p)$ from Eq.~(\ref{alternative}) yields, in the leading order, Eq.~(\ref{phiwkb2}).  As it is clear from Eq.~(\ref{alternative}), the WKB solution breaks down in a vicinity of the point $p=R_0^{-1}$, where the slow and fast WKB modes become strongly coupled. Here the quasi-stationarity does not hold.

We now proceed, therefore,  to the non-quasi-stationary region $-1\leq p\lesssim p_f$ (a more restrictive condition will appear \textit{a posteriori}). It is easier to deal with it in terms of $u(p)$, rather than $\phi(p)$. Here we can treat the $E\phi$ term in
Eq.~(\ref{ode2}) perturbatively:
$\phi(p)=1+\delta\phi(p)$, where
$\delta\phi\ll 1$ \cite{Assaf1,Assaf2}. As a result,
Eq.~(\ref{ode2}) becomes an inhomogeneous first-order
equation for $u(p)=\delta \phi^{\prime}(p)$:
\begin{equation}\label{odebulk}
\frac{1}{2N}(1-p^2)u^{\prime}+(p-1)\left(p-\frac{1}{R_0}\right)u=-E\,,
\end{equation}
which can be solved by variation of parameter. For two-body reactions the corresponding \textit{homogeneous} equation, which coincides with the quasi-stationary equation (\ref{ode2wkb}), is exactly solvable.  As a result,
one can solve Eq.~(\ref{odebulk}) in the entire non-quasi-stationary region which includes both $p<p_f$ and  $|p-p_f|\ll p_f$. The solution
is
\begin{eqnarray}
u(p)=-2NE e^{2N\left[p-\left(1+\frac{1}{R_0}\right)\ln(1+p)\right]}\int_{-1}^{p}\frac{\exp\left\{2N\left[s-\left(1+\frac{1}{R_0}\right)\ln
(1+s)\right]\right\}}{1-s^2}ds\,, \label{ubulk}
\end{eqnarray}
where the arbitrary constant is chosen so as to obey the boundary condition $u(-1)\simeq 0$. Note, that the integrand in Eq.~(\ref{ubulk}) is regular at $s=-1$, so the perturbative solution is well-behaved.  Solution (\ref{ubulk}) remains valid as long as
$\phi$ is close to $1$. As one can check, this holds for $1-p\gg N^{-1/2}$, \textit{cf}. Refs.~\cite{Assaf1,Assaf2}. The
perturbative solution (\ref{ubulk}) can be matched with the quasi-stationary solution (\ref{phiwkb2}), \textit{e.g.} at $N^{-1/2}\ll pR_0-1 \ll 1$ \cite{matching}.

Solution (\ref{ubulk}) simplifies in the ``left region" $p<p_f$, not too close to $p_f$. By Taylor-expanding the integrand in Eq.~(\ref{ubulk})
(which is a monotone increasing function of $p$ for $p<p_f$)
in the vicinity of $s=p$, we obtain
\begin{equation}\label{bulkleft}
u(p)^{left}\simeq -\frac{E}{(p-1)(p-R_0^{-1})}\,.
\end{equation}
This result (which holds in the region $1-pR_0\gg N^{-1/2}$) has a simple meaning: here the first-derivative term in Eq.~(\ref{odebulk}) is negligible.  To neglect this term in Eq.~(\ref{odebulk}) [or the term proportional to $\phi^{\prime\prime}(p)$ in Eq.~(\ref{ode2})] is the same as to disregard the two-body reaction $2A\to \emptyset$ compared with the one-body reactions of branching and decay. This is indeed a legitimate approximation at small $n$ \cite{kessler,Assaf4}. Note that, not too close to $p=p_f$, $u(p)^{left}$ is exponentially small in $N$, so that $\phi \simeq 1$ up to an exponentially small correction. Putting $\phi=1$ in the left region, however,  would be too a crude approximation, as it would only give a trivial left tail of the QSD: $\pi_1=\pi_2=\dots=0$ \cite{leftQSD}. Correspondingly, the solution in the left region cannot be obtained from the WKB approximation.

We can now find the eigenvalue $E$ by matching  the quasi-stationary solution (\ref{phiwkb2}) and the perturbative non-quasi-stationary solution (\ref{ubulk}) in their joint validity region $N^{-1/2}\ll pR_0-1 \ll 1$ \cite{matching}. For $pR_0-1\gg N^{-1/2}$, the integral
in Eq.~(\ref{ubulk}) can be evaluated by the saddle point
approximation. The saddle point is at $p=p_f=R_0^{-1}$, and the result is
\begin{eqnarray}
u(p)\simeq-\frac{2E\sqrt{\pi N}R_0^{3/2}}{\sqrt{R_0+1}(R_0-1)}e^{-2N\left[\frac{1}{R_0}-\left(1+\frac{1}{R_0}\right)\ln
\left(1+\frac{1}{R_0}\right)\right]+2N\left[p-\left(1+\frac{1}{R_0}\right)\ln
(1+p)\right]}\,. \label{ubulkmatch}
\end{eqnarray}
Matching this result with the quasi-stationary solution (\ref{phiwkb2}), we
find
\begin{equation}\label{e12}
E=\sqrt{\frac{N(R_0+1)}{4\pi}}\frac{(R_0-1)^2}{R_0^{5/2}}e^{-NS_0}\,,
\end{equation}
where
\begin{equation}\label{S0}
S_0= 2\left[1-\ln2 -\frac{1+\ln 2}{R_0}
+\left(1+\frac{1}{R_0}\right)\,\ln\left(1+\frac{1}{R_0}\right)
\right].
\end{equation}
The MTE, in physical units, is $\tau_{ex}=(\lambda E)^{-1}$ with $E$ from Eq.~(\ref{e12}), in agreement with Ref. \cite{kessler}. As $R_0\to \infty$ the decay reaction
$A \to 0$ becomes irrelevant, and one recovers the result for the branching-annihilation model \cite{Assaf2,kessler,turner}.  When $R_0-1\ll 1$, the system is close to the transcritical bifurcation of the deterministic rate equation. Here the Fokker-Planck approximation to the master equation is applicable \cite{Doering,kessler}. The corresponding asymptote of Eq.~(\ref{e12}),
$$
E=\sqrt{\frac{N}{2 \pi}}\,(R_0-1)^2\,e^{-\frac{N}{2} (R_0-1)^2}\,,
$$
is valid when $R_0-1\gg N^{-1/2}$, so that $E$ is still exponentially small in $N$.

\begin{figure}
\includegraphics[width=9.0cm,height=6.6cm,clip=]{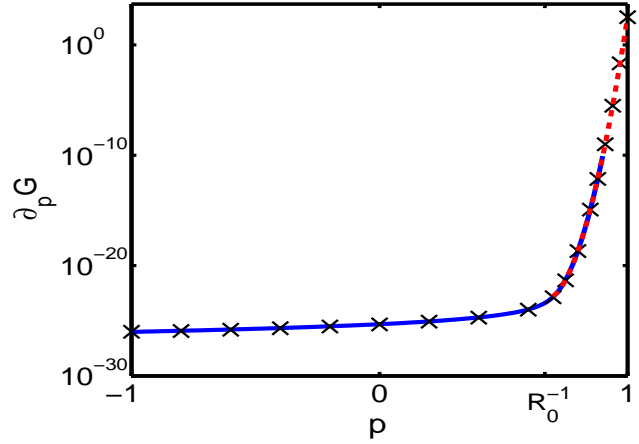}
\caption{Branching-annihilation-decay. Shown is the $p$-derivative $\partial_p G$ of the probability generating function $G$
at $t_r\ll t\ll 1/E$ for $N=10^3$ and $R_0=1.5$. The solid line is the absolute value of the perturbative solution~(\ref{ubulk}); the dashed line is the absolute value of the quasi-stationary
solution~(\ref{phiwkb2}). In the joint
region of their validity the two lines are indistinguishable. The crosses indicate the values obtained by a numerical solution of Eq.~(\ref{pde2}) for $n_0=20$
particles at $t=0$ and boundary conditions $G(1,t)=1$ and $\partial_p G(-1,t)=0$.} \label{genfun}
\end{figure}

Having found $E$, we have a complete solution for $u(p)$, given by Eqs.~(\ref{phiwkb2}) and (\ref{ubulk}). Now one can find the QSD by using Eq.~(\ref{prob}) for $n={\cal O}(1)$ and Eq.~(\ref{cauchy}) for $n\gg 1$. The results coincide with those obtained by Kessler and Shnerb  \cite{kessler} by the ``real-space" WKB  approximation, so we will not present them here. The large-$n$ tail of the QSD  decays faster than exponentially, thus justifying our \textit{a priori} assumption that $\phi(p)$ is an entire function in the complex $p$-plane. Shown in Fig.~\ref{genfun} is a comparison between the analytical and numerical solutions for $\partial_p G \simeq -u(p) e^{-Et}$ at a time $t_r\ll t\ll 1/E$, when  $\partial_p G\simeq -u(p)$.

\subsection{Branching and triple annihilation}\label{sec3}
Here we again consider a metastable population on the way to extinction, but now a three-body reaction is present.  Our model system includes two reactions: the branching
$A\to\hspace{-4.3mm}^{\lambda}\hspace{2mm}2A$ and the triple annihilation $3A\to\hspace{-4.3mm}^{\mu}\hspace{2mm} \emptyset$. The deterministic rate equation,
\begin{equation}
\label{rate3}
\dot{\bar{n}}=\lambda \bar{n}-\frac{\mu}{2}\bar{n}^3,
\end{equation}
has two relevant fixed points: the repelling point $n=0$ and the attracting point
$n_1=(2\lambda/\mu)^{1/2}\equiv N\gg 1$. According to Eq.~(\ref{rate3}),
the system size approaches $\bar{n}=n_1$ after the relaxation time
$t_r=\lambda^{-1}$, and stays there forever. Contrary to this prediction, fluctuations drive the population to extinction. Upon rescaling time $t\to \lambda t$, the master equation reads
\begin{eqnarray}
\frac{dP_n(t)}{dt}=(n-1)P_{n-1}-n P_n +\frac{1}{3 N^2}\left[(n+3)(n+2)(n+1)P_{n+3}-n(n-1)(n-2)P_n\right],\nonumber\\\label{master3}
\end{eqnarray}
whereas the evolution equation for $G(p,t)$ is
\begin{equation}\label{pde3}
\frac{\partial G}{\partial t}=\frac{1}{3N^2}(1-p^3)\frac{\partial^3
G}{\partial p^3}+p(p-1)\frac{\partial G}{\partial p}\,.
\end{equation}
At $t\gg t_r$, Eq.~(\ref{G2}) holds, and the ordinary differential equation for the lowest excited eigenfunction $\phi(p)$ is
\begin{equation}\label{ode3}
\frac{1}{3N^2}(1-p^3)\phi^{\prime\prime\prime}+p(p-1)\phi^{\prime}+E\phi=0\,.
\end{equation}
This equation has three singular points in the complex $p$-plane. These are the roots of $1-p^3$: one real, $p_1=1$, and two complex,
$p_2=e^{2\pi i/3}$ and
$p_3=e^{4\pi i/3}$. Since $\phi(p)$ must be analytical in all these points, $\phi(p)$ must satisfy three conditions:
\begin{equation}\label{boundcon}
    p_i (p_i-1)\phi^{\prime}(p_i)+E\phi(p_i)=0\,\;\;\;i=1,2,3.
\end{equation}
Here the $p$-derivative is in the complex plane. For $i=1$ Eq.~(\ref{boundcon}) yields $\phi(p=1)=0$. As $E$ turns out to be exponentially small in $N\gg 1$, we can neglect small terms proportional to $E$ in the conditions for $i=2$ and $3$ and obtain $\phi^{\prime}(p=e^{2\pi i/3})\simeq 0$ and $\phi^{\prime}(p=e^{4\pi i/3})\simeq 0$.

\begin{figure}
\includegraphics[width=9.0cm,height=6.6cm,clip=]{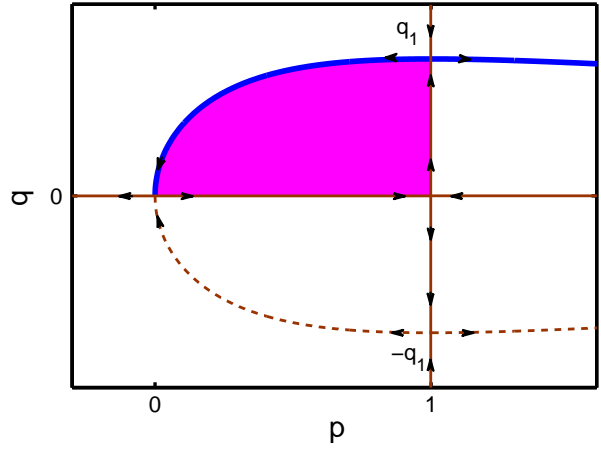}
\caption{Branching and triple annihilation. Shown are zero-energy lines of Hamiltonian (\ref{Ham2}) on the $(p,q)$ phase plane. The
thick solid line is the instanton $q=-S^{\prime}(p)=\psi(p)$, given by
Eq.~(\ref{sprime}). Here $q_1=1$, and the area of the shaded region is equal to
$S_0$ from Eq.~(\ref{s0}). The dashed line denotes a non-physical orbit.} \label{phase3}
\end{figure}

In the quasi-stationary region (the exact location of which will be determined later) Eq.~(\ref{ode3}) becomes
\begin{equation}\label{ode30}
\frac{1}{3N^2}(1-p^3)\phi^{\prime\prime\prime}+p(p-1)\phi^{\prime}=0\,.
\end{equation}
This equation is of second order for $u(p)=\phi^{\prime}(p)$, but it is not exactly solvable in terms of known special functions, and this is a typical situation for three-body, four-body, $\dots$, reactions. The presence of the large parameter $N\gg 1$ justifies the WKB ansatz $\phi(p)=a(p) e^{-N S(p)}$. It yields, in the leading order of $N\gg1$, a stationary Hamilton-Jacobi equation $H[p,-S^{\prime}(p)]=0$ with Hamiltonian
\begin{equation}\label{Ham2}
    H(p,q)=\left[p-\frac{(1+p+p^2)q^2}{3}\right]q(p-1).
\end{equation}
Here again, in addition to the trivial zero-energy lines $q=0$ and $p=1$, one obtains an instanton orbit
\begin{equation}\label{sprime}
    q=\psi(p)\equiv \left(\frac{3p}{1+p+p^2}\right)^{1/2}
\end{equation}
which connects the fixed points $(1,q_1=n_1/N=1)$ and $(0,0)$ in the $(p,q)$ plane, see Fig.~\ref{phase3}. The instanton corresponds to the fast-mode WKB solution, whereas the orbit $q=0$ corresponds to the slow-mode WKB solution, similarly to the previous example.

Again, it is simpler to do the actual calculations for $u(p)=\phi^{\prime}(p)$, rather than for $\phi(p)$. Using the WKB ansatz $u(p)=b(p)e^{-N S(p)}$ in the quasi-stationary equation
\begin{equation}\label{u3wkb}
\frac{1}{3N^2}(1+p+p^2)u^{\prime\prime}-p u=0\,,
\end{equation}
we obtain
\begin{equation}
\frac{(1+p+p^2)}{3N^2}[N^2(S^{\prime})^2
b-2NS^{\prime}b^{\prime}-N S^{\prime\prime}b]-pb=0,\label{pert3}
\end{equation}
where we have neglected the sub-subleading term proportional to $b^{\prime\prime}/N^2$. In the leading order we obtain
$S^{\prime}(p)=-\psi(p)$ [the solution with $S^{\prime}(p)=\psi(p)$ is non-physical and must be discarded].  The arbitrary constant can be fixed by putting $S(1)=0$, and we obtain
\begin{equation}\label{action3}
S(p)=-\int_{1}^{p}\psi(x)dx,
\end{equation}
with $\psi(x)$ from Eq.~(\ref{sprime}). This result can be expressed via elliptic integrals, but we will not need these cumbersome formulas.

In the subleading order Eqs.~(\ref{pert3}) and
(\ref{action3}) yield a first-order ordinary differential equation for $b(p)$
whose general solution is
\begin{equation}
b(p)=\frac{C(1+p+p^2)^{1/4}}{p^{1/4}}\,.
\end{equation}
We demand $u(1)\simeq -n_1=-N$ [see Eq.~(\ref{dGdp})] and obtain the quasi-stationary WKB solution for $u(p)$:
\begin{equation}\label{phiwkb3}
u^{WKB} (p)=-\frac{N(1+p+p^2)^{1/4}}{(3p)^{1/4}}e^{-NS(p)}
\end{equation}
with $S(p)$ from Eq.~(\ref{action3}). Now one can check that asymptote~(\ref{phiwkb3}) is valid
when $p\gg N^{-2/3}$; otherwise it is not justified to neglect the term $b^{\prime\prime}(p)$ in Eq.~(\ref{pert3}). Again, the quasi-stationarity and the WKB approximation break down in a vicinity of the point where the fast and slow WKB modes are strongly coupled. In this example this point is at $p=0$,  whereas in the previous example it was at $p=p_f\neq 0$. That the WKB breaks down here at $p=0$ is a special, non-generic situation resulting from the absence of the linear decay process $A\to 0$ from the set of reactions $A\to 2A$ and $3A\to 0$ we are dealing with.

To remedy the divergence of the WKB solution at $p=0$, one need to account for a deviation from quasi-stationarity. The corresponding non-quasi-stationary solution of Eq.~(\ref{ode3}) is perturbative in $E$, as in the previous example, so the equation we need to solve is
\begin{equation}\label{pert10}
\frac{1}{3N^2}(1-p^3)u^{\prime\prime}+p(p-1)u=-E\,.
\end{equation}
The corresponding homogeneous equation, Eq.~(\ref{ode30}), is not solvable in known special functions. Therefore, we will solve Eq.~(\ref{pert10})
approximately in two separate regions and match the solutions in their joint region of validity.

The first region, which we call ``left",  is $p<0$ (and not too close to zero, see below).
Here we can neglect the $u^{\prime\prime}$-term in Eq.~(\ref{pert10}) and obtain
\begin{equation}
u^{left}(p)\simeq \frac{E}{p(1-p)}\,.
\label{left}
\end{equation}
This asymptote, valid when $-p\gg N^{-2/3}$, corresponds to neglecting the high-order reaction $3A \to 0$ at small population sizes. As in the previous example, $u^{left}(p)$ is exponentially small. By choosing an exponentially small solution for $u(p)$ in the left region, we effectively discarded two other linearly independent solutions of Eq.~(\ref{ode30}) which are singular at $p=e^{2\pi i/3}$ and $p=e^{4\pi i/3}$.
As $p_f=0$ here, one can actually put $u=0$ in the left region and still accurately determine the QSD \cite{leftQSD}.

The second region is the boundary layer $|p|\ll 1$, where Eq.~(\ref{pert10}) becomes
\begin{equation}
\label{oder3}
\frac{1}{3N^2}u^{\prime\prime}-pu=-E\,,
\end{equation}
The general solution of this equation is
\begin{eqnarray}
u^{bl} (p)=\left[c_1+\alpha^2 \pi E \int_0^pBi(\alpha s)ds\right] Ai(\alpha p)+\left[c_2-\alpha^2 \pi E \int_0^pAi(\alpha
s)ds\right] Bi(\alpha p)\,,\label{phibl3u}
\end{eqnarray}
where $Ai(y)$ and $Bi(y)$ are the Airy functions of the first and
second kind, respectively \cite{Abramowitz}, and $\alpha=(3N^2)^{1/3}$.

\begin{figure}
\includegraphics[width=9.0cm,height=6.6cm,clip=]{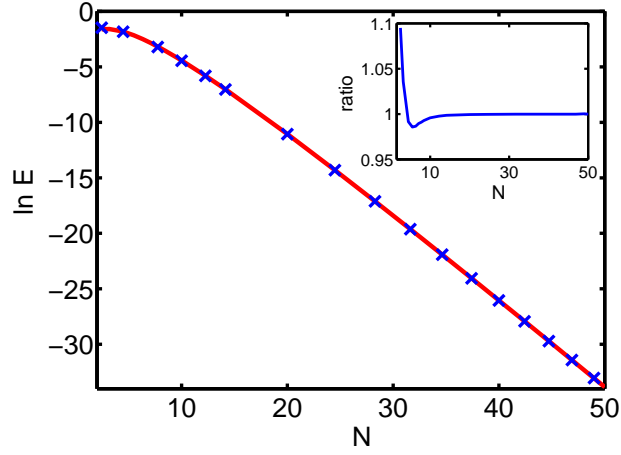}
\caption{Branching and triple annihilation. Shown is a comparison between the extinction rate (\ref{exttime3}) (solid line) and the extinction rate $-[\ln(1-P_0^{num}(t))]/t$ (crosses) found from a
numerical solution of the master
equation (\ref{master3})  at
different $N$. The inset shows the ratio of the two rates.} \label{extrate}
\end{figure}

Now we can find the unknown constants $c_1$ and $c_2$ (assuming for a moment that $E$ is known) by matching the asymptotes (\ref{left}) and (\ref{phibl3u}) in their common region
$N^{-2/3}\ll -p \ll 1 $. As $u^{left}(p)$ is exponentially small at $N^2|p|^3\gg 1$, the boundary layer solution $u^{bl}(p)$ from Eq.~(\ref{phibl3u}) must also be exponentially small there. Evaluating the integrals in Eq.~(\ref{phibl3u}) at $p=-\infty$ and using the identities
$\int_{-\infty}^0 Bi(s)ds=0$ and $\int_{-\infty}^0 Ai(s)ds=2/3$,
we arrive at
\begin{equation}\label{coeff3}
c_1\simeq 0\;,\;\;\;c_2\simeq -\frac{2\pi E N^{2/3}}{3^{2/3}}\,.
\end{equation}
Now we can find the extinction rate $E$ by matching the asymptotes of
$u^{WKB}(p)$ and $u^{bl}(p)$ in their common region $N^{-2/3}\ll p\ll 1$.
The $p \ll 1$ asymptote of the WKB solution (\ref{phiwkb3}) is
\begin{equation}\label{wkbapprox3}
u^{WKB}\simeq
-\frac{N}{(3p)^{1/4}}e^{-NS_0}e^{(2/\sqrt{3})Np^{3/2}},
\end{equation}
where
\begin{eqnarray}
S_0=\int_{0}^{1}\left(\frac{3x}{1+x+x^2}\right)^{1/2}dx=0.836367\dots,\label{s0}
\end{eqnarray}
is the shaded area in Fig.~\ref{phase3}. Let us obtain the $p\gg N^{-2/3}$
asymptote of $u^{bl}(p)$ (\ref{phibl3u}). First, for $z\gg 1$
\cite{Abramowitz}
\begin{equation}
\label{largearg}
Ai(z)\simeq \frac{e^{-(2/3)z^{3/2}}}{2\pi^{1/2} z^{1/4}}\;,\;\;Bi(z)\simeq \frac{e^{(2/3)z^{3/2}}}{\pi^{1/2} z^{1/4}}\,.
\end{equation}
Now we need to evaluate the integrals in Eq.~(\ref{phibl3u}). As we are interested in the region of $N^2p^3\gg 1$, the integral of $Ai(\alpha s)$ can be evaluated by putting $p=\infty$ and using the saddle point approximation, arriving at
$$\int_0^{\infty} Ai\left[(3N^2)^{1/3}s\right]ds=\frac{1}{3^{4/3}N^{2/3}}\,.$$
The main contribution to the integral of $Bi(\alpha s)$ at $N^2p^3\gg 1$ comes from a vicinity of $s=p$, where  $Bi(\alpha s)$ is exponentially large, see Eq.~(\ref{largearg}). Expanding the exponent in a Taylor series around $s=p$, we obtain in the leading order
$$\int_0^{p} Bi
\left[(3N^2)^{1/3}s\right]ds\simeq\frac{e^{(2/\sqrt{3})Np^{3/2}}}
{3^{7/12}\pi^{1/2} N^{7/6}p^{3/4}}\,.$$
Now one can see from Eq.~(\ref{coeff3}) that the main contribution to $u^{bl}(p)$ (\ref{phibl3u}) comes from the $Bi(\alpha p)$ term, and we obtain
\begin{equation}\label{uapprox}
u^{bl}\simeq-\frac{3^{1/4}\sqrt{\pi N}E}{p^{1/4}}
e^{(2/\sqrt{3})Np^{3/2}}\,.
\end{equation}
Matching Eqs.~(\ref{wkbapprox3}) and (\ref{uapprox}), we obtain
\begin{equation}\label{exttime3}
E=\sqrt{\frac{N}{3\pi}}e^{-NS_0}\,.
\end{equation}
The MTE  in physical units is given by $\tau_{ex}=(\lambda E)^{-1}$,
which is exponentially large in $N$, as expected. A comparison between the analytical result for the extinction rate (\ref{exttime3}) and a
numerical result, obtained by solving (a truncated version of) master equation (\ref{master3}), is shown in Fig.~\ref{extrate}.
For $N\gg 1$ the agreement is excellent.

Now let us calculate the QSD. Combining Eq.~(\ref{cauchy}) with Eqs.~(\ref{G2}) and  (\ref{qsd}), we obtain
\begin{equation}\label{exact}
    \pi_{n\geq 1}=-\frac{1}{2\pi n
i}\oint\frac{u(p)}{p^n}dp\,.
\end{equation}
For $n\gg 1$ we can use the WKB asymptote~(\ref{phiwkb3}):
\begin{eqnarray}
\pi_{n\gg1}\simeq-\frac{1}{2\pi n i}\oint\frac{u^{WKB}(p)}{p^n}dp =\frac{N}{2\pi n i}\oint
\frac{(1+p+p^2)^{1/4}}{(3p)^{1/4}}\frac{\exp\left[N \int_{1}^{p}\psi(x)dx\right]}{p^{n}}\,dp,
\end{eqnarray}
with $\psi(x)$ given by Eq.~(\ref{sprime}). As $N\gg 1$ and $n\gg 1$, this integral can be
evaluated via saddle point approximation \cite{orszag}. Let us denote
$f(p)=N\int_{1}^{p}\psi(x)dx-n\ln p$. The saddle point equation $f^{\prime}(p_*)=0$ reduces to a cubic equation
\begin{equation}\label{saddle}
\frac{3 p^3}{1+p+p^2}=\left(\frac{n}{N}\right)^2,
\end{equation}
which has one and only one real root  $p_*=p_*(n/N)$.  As $f^{\prime\prime}(p_*)>0$, we must choose a
contour in the complex $p$-plane which goes through this root perpendicularly to the
real axis. The Gaussian integration yields
\begin{equation}\label{wkb3b}
\pi_{n\gg1}=\frac{N(1+p_*+p_*^2)^{1/4}}{n \sqrt{2\pi
f^{\prime\prime}(p_*)}\,(3p_*)^{1/4}}\frac{\exp\left[N
\int_{1}^{p_*}\psi(x)dx\right]}{p_*^{n}};
\end{equation}
we omit a cumbersome expression for $f^{\prime\prime}(p)$. Note, that for $n\gg 1$, the saddle point $p_*$ is always obtained in the region where $u^{WKB}(p)$ is valid, see below. Let us calculate the $1\ll n\ll N$ and $n\gg N$ asymptotes of Eq.~(\ref{wkb3b}) with an exponential accuracy,
$\ln \pi_n \simeq f(p_*)$. For $n\ll N$ the saddle point,
given by Eq.~(\ref{saddle}), is obtained at $p_*=[n/(\sqrt{3}N)]^{2/3}\ll 1$. Here it suffices, in the leading order in $n/N$, to put $p_*=0$ in the upper bound of the integral in Eq.~(\ref{action3}). Then the integral yields $S_0$ from Eq.~(\ref{s0}).  For $n\gg N$ we obtain $p_*=[n/(\sqrt{3}N)]^2\gg 1$. Here a dominant contribution to the integral in Eq.~(\ref{action3}) comes from the region of $p_*\gg 1$ which enables one to simplify the integrand. The resulting asymptotes are
\begin{eqnarray}\label{probleftright}
\ln \pi_n&\simeq& N\left[-S_0+\frac{2n}{3N}\ln \frac{N}{n}+{\cal
O}\left(\frac{n}{N}\right)\right],\,\;n\ll N \nonumber\\
\ln \pi_n&\simeq& N\left[-\frac{2n}{N}\left(\ln\frac{n}{N}-1-\ln
\sqrt{3}\right)+{\cal O}(1)\right],\,\;n\gg N\,.\nonumber\\
\end{eqnarray}
Notice that each of these tails of the QSD are non-Gaussian. The $n\gg N$ tail decays faster than exponentially, thus justifying \textit{a posteriori} our assumption that $\phi(p)$ is an entire function in the complex $p$-plane.

At $|n-N|\ll N$ the saddle point, given by Eq.~(\ref{saddle}), is obtained at
$p_*=1+(n-N)/N$, and we arrive at a Gaussian asymptote
\begin{equation}\label{gauss3}
\pi_n\simeq \frac{1}{\sqrt{2\pi N}}e^{-(n-N)^2/(2N)};
\end{equation}
the preexponent is fixed by normalization. Equation~(\ref{gauss3}) holds for $|n-N|\ll N^{2/3}$; this condition is tighter than $|n-N|\ll N$. Note, that the Gaussian asymptote of the QSD can also be found by directly calculating the mean and variance of the distribution. These (and other higher cumulants of the distribution) can be found by using derivatives of $G(p,t)$ with respect to $p$ at $p=1$, see Eq.~(\ref{genprob}). Indeed, from Eqs.~(\ref{genprob}) and (\ref{genfun}), the mean of the QSD (at times $t_r\ll t\ll \tau_{ex}$) is given by $\bar{n}=\partial_p
G|_{p=1}\simeq -u(p=1)=N$, where here we have used $u=u^{WKB}(p)$ given by Eq.~(\ref{phiwkb3}). In its turn the variance in the leading order is
\begin{eqnarray}
V=\bar{n^2}-\bar{n}^2=\sum_{n=0}^{\infty}n^2P_n(t)-\left(\sum_{n=0}^{\infty}nP_n(t)\right)^2 =\left.\left[\partial_{pp} G+\partial_p G -(\partial_p G)^2\right]\right|_{p=1} \simeq -u^{\prime}(1)-u(1)-[u(1)]^2\simeq N\nonumber\,,
\end{eqnarray}
recovering the Gaussian asymptote (\ref{gauss3}).

\begin{figure}
\includegraphics[width=9.0cm,height=6.6cm,clip=]{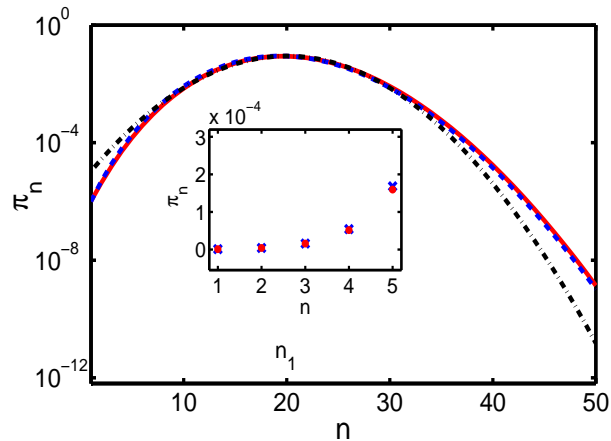}
\caption{Branching and triple annihilation. Shown is the natural logarithm of the QSD versus $n$ for $N=20$. The dashed line is
WKB solution (\ref{wkb3b}), the dash-dotted line is the Gaussian
approximation (\ref{gauss3}), and the solid line is the numerical
solution of the (truncated) master equation
(\ref{master3}). Inset: the $n\ll N$ asymptote of the QSD obtained analytically [Eqs.~(\ref{pi123}) and (\ref{smalln3})] ($\times$'s) and numerically (fat dots).} \label{prob3b}
\end{figure}

At $n={\cal O}(1)$ the QSD can be evaluated directly from
$$
\pi_n=-\left.\frac{1}{n!}\frac{d^{n-1} u(p)}{d p^{n-1}}\right|_{p=0},\;\;\;\;\;\; n\geq1.
$$
Here one should use the boundary-layer solution around $p=0$, given by Eqs.~(\ref{phibl3u}) and (\ref{coeff3}). This yields
\begin{equation}\label{pi123}
\pi_1=\frac{\Gamma(1/3) E N^{2/3}}{3^{1/3}}\;,\;\;\pi_2=\frac{\pi EN^{4/3}}{3^{1/6}\Gamma(1/3)}\;,\;\;\mbox{and}\;\;
\pi_3=\frac{E N^2}{2}.
\end{equation}
To calculate other $n={\cal O}(1)$ terms, one can use a recursion relation obtainable from the master equation (\ref{master3}) with $\dot{P}_n=0$.  Indeed, at  $n\ll N$ one can  neglect
the terms $n(n-1)(n-2)P_n$ and $(n-1)P_{n-1}$ compared with the terms $(n+3)(n+2)(n+1)P_{n+3}$ and $nP_n$, respectively, and arrive at the following relation:
\begin{equation}\label{smalln3}
\pi_{n+3}=\frac{3N^2 n }{(n+3)(n+2)(n+1)}\pi_n\,.
\end{equation}
Note, that the small-$n$ (\ref{smalln3}) and the WKB (\ref{wkb3b}) segments of the QSD have a joint region of validity at $1\ll n\ll N$.

A comparison between WKB result (\ref{wkb3b}) and a numerical solution of (a truncated version of) master equation (\ref{master3}) is shown in
Fig.~\ref{prob3b}. The inset compares the $n \ll N$ analytical asymptote [see Eqs.~(\ref{pi123}) and (\ref{smalln3})] with numerical results. Excellent agreement is observed in both cases. It can be also seen that the Gaussian approximation (\ref{gauss3}) strongly overestimates the QSD in the low-$n$ region, and underestimates it in the high-$n$ region.

\section{Discussion}\label{discussion}
The $p$-space representation renders a unique perspective to theory of large fluctuations in populations undergoing Markovian stochastic gain-loss processes.  The stationary distribution of the population size is encoded in the ground-state eigenfunction of a Sturm-Liouville (spectral) problem for the probability generating function. In the case of a long-lived metastable population on the way to extinction, the MTE and the quasi-stationary distribution of population size are encoded in the eigenfunction of the lowest excited state. The uniqueness of solution in these problems is guaranteed by the condition that the probability generating function is an entire function on the whole complex $p$-plane except at infinity. As this work has demonstrated (see also Refs. \cite{Assaf2,Assaf1,Assaf,Kamenev1,Kamenev2,AKM}), the $p$-space representation in conjunction with the WKB approximation and other perturbation tools
employing a large parameter $N\gg 1$ (the mean population size in the stationary or metastable state) yields accurate results for extreme statistics in a broad class of problems of stochastic population dynamics. Such an accuracy is usually impossible to attain via the van Kampen system size expansion which approximates the exact master equation by a  Fokker-Planck equation.

How does the $p$-space approach compare with the ``real" space WKB method of Refs. \cite{dykman,kessler,MS,EK,Assaf4} when the stationary or metastable population size is large, $N\gg 1$? One advantage of the $p$-space representation is that, for two-body reactions, there is no need in  the WKB approximation, as the quasi-stationary equation in this case is always solvable exactly. Another advantage appears when the WKB solution for $G(p,t)$ is valid for every $p\gtrsim 0$, as occurs in the molecular hydrogen production problem, Sec. III.
In such cases one directly finds the entire probability distribution function, including the region of
small $n={\cal O}(1)$.  In the real-space approach
a separate (non-WKB) treatment of the $n={\cal O}(1)$ region, and a matching with the WKB-solution valid at $n\gg 1$ would be needed \cite{Assaf4}.

Still, from our experience, every problem which includes the large parameter $N\gg 1$, and can be solved
in the $p$-space, can be also solved in the ``real" space. Furthermore, for populations exhibiting escape to infinity \cite{MS}, escape to another metastable state \cite{EK}, or Scenario B of extinction \cite{Assaf4}, the $p$-space representation meets significant difficulties. One difficulty is that one should account for a constant-current WKB solution in these cases \cite{MS,EK,Assaf4}. The constant-current solution comes from the deterministic line  $p=1$ of the phase plane of the underlying classical Hamiltonian. In the $p$-representation this line is vertical, as in Figs. \ref{phasehyd}, \ref{phasedb} and \ref{phase3},  and so the constant-current solution cannot be easily accounted for. In addition, it is unclear how to deal with the region of non-uniqueness of $q=q(p)$ which is inherent, in the $p$-representation, in these cases. There are two WKB solutions in this region, one of them exponentially small compared with the other. The real-space approach avoids these difficulties, and the solution in these cases can be worked out in a straightforward manner \cite{MS,EK,Assaf4}.

An important advantage of the $p$-space representation stems from the fact that the evolution equation for $G(p,t)$ is \textit{exactly} equivalent to the original master equation. Therefore, the $p$-space approach is especially valuable for exact analysis, as illustrated by the example of molecular hydrogen production, see Ref. \cite{green} and Section III.

Finally, generalization of the $p$-representation to interacting multi-species populations is quite straightforward, see Ref.~\cite{KM}. The resulting multi-dimensional evolution equation for the probability generating function can be analyzed in WKB approximation. As of present only the leading-order WKB-approximation for population extinction is available, and this is regardless of whether one uses the $p$- or $n$-space approach. In the leading WKB order the problem again reduces to finding a nontrivial zero-energy trajectory of the corresponding classical Hamiltonian, and the action along this special trajectory. This problem can be solved numerically. If additional small parameters are present, the problem may become solvable analytically, again in both $p$- and $n$-spaces~\cite{KM,DSL,MS2}.

\section*{Acknowledgments}
We are grateful to Alex Kamenev for fruitful discussions. This work was supported by the Israel Science Foundation (Grant No. 408/08).


\begin{thebibliography}{99}
\bibitem{matching} The joint region of (\ref{phiwkb2}) and (\ref{ubulk}) is actually broader and spans from $pR_0-1\gg N^{-1/2}$ to $1-p\gg
N^{-1/2}$.
\bibitem{leftQSD} The solution for $u(p)$ in the left region $p<p_f$ determines the left tail of the QSD in the generic case where $p_f>0$. If,
however, $p_f=0$, the left-region solution is not needed for the purpose of the QSD and can be approximately put to zero.
\end{thebibliography}
\end{document}